\documentstyle[version2,aps,epsf,titlepage]{revtex}
\draft
\begin{document}
\newcommand {\ua} {\uparrow}
\newcommand {\da} {\downarrow}
\newcommand {\sqt} {\frac{1}{\sqrt{2}}}
\twocolumn[{
\begin{title}
\begin{center}
Magnetic Impurity in a Metal with Correlated Conduction Electrons:\\ 
An Infinite Dimensions Approach
\end{center}
\end{title}
\author{Benny Davidovich$^1$ and V. Zevin$^{1,2}$}
\begin{instit}
$^1$The Racah Institute of Physics, The Hebrew University of Jerusalem, 
91904 Jerusalem, Israel \\
$^2$Max-Planck-Institut f\"ur Physik Komplexer Systeme,
Au{\ss}enstelle Stuttgart,\\ Postfach 80 06 65, 70506 Stuttgart, Germany
\end{instit}
\begin{abstract}
We consider the Hubbard model with a magnetic Anderson impurity coupled to a 
lattice site. In the case of infinite dimensions, one-particle correlations 
of the impurity electron are described by the effective Hamiltonian of the 
two-impurity system. One of the impurities interacts with a bath of 
free electrons and represents the Hubbard lattice, and the other is coupled 
to the first impurity by the bare hybridization interaction. A study of the 
effective two-impurity Hamiltonian in the frame of the 1/N expansion and 
for the case of a weak conduction-electron interaction ( small $U$ ) reveals 
an enhancement of the usual exponential Kondo scale. However, an 
intermediate interaction  ( $U/D = 1 - 3$ ), treated by the variational 
principle, leads to the loss of the exponential scale. The Kondo temperature 
$T_K$ of the effective two-impurity system is calculated as a function of the 
hybridization parameter and it is shown that $T_K$ decreases with an increase 
of $U$.
The non-Fermi-liquid character of the Kondo effect in the intermediate 
regime at the half filling is discussed. 
\end{abstract}
}]

\narrowtext
\section{Introduction}
The physics
of a magnetic impurity embedded in a metal with non-interacting conduction 
electrons is well now understood and documented 
\cite{TsvelickWiegmann,HewsonBook}. Its characteristic feature 
is the many-body non-perturbative renormalization of the 
impurity-host interaction which turns to be strong at temperatures lower 
than the Kondo temperature $T_K$. This $T_K$ is an exponential in the 
inverse impurity-host coupling parameter. Recently both the Kondo-spin model 
\cite{LeeToner,FurusakiNagaosa,Avi} and the Anderson impurity model were 
studied \cite{Li,Phillips,Avi2} in one-dimensional systems with interacting 
conduction electrons. It turns out that the influence of the 
conduction-electron interaction ( CEI ) on the Kondo effect 
\cite{footnote1} in a Luttinger liquidis is not trivial: the Kondo screening 
of the impurity spin may also be possible for a ferromagnetic exchange 
coupling $J$ \cite{FurusakiNagaosa,Avi}; for a sufficiently large CEI, a 
power-law dependence of $T_K$ upon $J$ replaces the usual exponential law 
\cite{LeeToner,FurusakiNagaosa,Avi}; in the case of the Anderson 
model a small CEI enhances $T_K$ ( just because more 
conduction electrons effectively participate in the interaction with an 
impurity ) but this trend changes for a large enough CEI and $T_K$ falls due 
to the supression of the charge transfer between the impurity and the 
lattice \cite{Li,Avi2}. A generalized Anderson impurity model for a 
Luttinger liquid which was introduced in \cite{Avi2} and studied there by 
the renormalization-group techniques shows a rich phase diagram in the 
parameter space.

The interest in the Kondo effect in the interacting host for 2d and 3d 
was sparked by the discovery \cite{Bruggeretal} of heavy-fermions in 
$Nd_{2-x}Ce_xCuO_4$ \cite{FZZ,FZ,TZZ}. The Schrieffer-Wolff 
transformation for the magnetic impurity coupled to the Hubbard host was 
discussed in \cite{FuldeSchork} and the Kondo-spin model in 
the above host was investigated in \cite{KhaliullinFulde}. 
The Anderson impurity model for interacting conduction electrons 
was considered too. Its ground state energy was calculated in the frame of 
the 1/N expansion \cite{Schork} and the Non-Crossing-Approximation( NCA ) 
theory was generalized to the interacting case \cite{TZZ2}. As was shown 
first in \cite{KhaliullinFulde}, two-particle Green's functions of host 
eletrons ( vertex corrections ) are an essential ingredient of the correct 
theory of the Kondo effect in the interacting host. Actual calculations were 
carried out in the lowest order in the Hubbard $U$ approximation  and an 
enhancement of $T_K$ \cite{footnote2} was obtained both 
for the Kondo-spin model \cite{KhaliullinFulde} and the Anderson  
\cite{Schork,TZZ2}. However, contrary to the 1d case,
the theory of the Anderson impurity  in 2d and 3d is not developed 
sufficiently beyond the case of the weak host interaction.

It is much easier to treat strongly interacting systems in the limit of 
infinite dimensions, $d \rightarrow \infty$  
\cite{MetznerVollhardt,MuellerHartmann,Vollhardt,LISAreview}. We use here 
the Local Impurity Self Consistent Approximation ( LISA ) 
\cite{GeorgesKotliar,Jarrell,LISAreview} to map the 
Hubbard host with an Anderson impurity to a simpler model. As we show, this
 system is a two-impurity system in which one of the impurities represents 
the Hubbard host \cite{LISAreview} and the other is the original ( bare ) 
impurity coupled solely to the first impurity by the bare hybridization 
interaction. In the LISA approach the effective two-impurity Hamiltonian 
preserves all the features of the one particle Green's function of the 
impurity electron. Using for simplicity the Bethe lattice in its large 
connectivity limit, we show by employing the NCA for a weak CEI  that the 
effective two-impurity Hamiltonian leads to an enhanced Kondo scale $T_K$ 
in full agreement with \cite{TZZ2,footnote2}. For an intermediate CEI we 
solve the effective two-impurity 
Hamiltonian by using the variational principle, and calculate numerically
the singlet and the triplet ground state energies. Their 
difference is no longer exponential in the inverse coupling parameter. 
For a small hybridization interaction it is linear in this parameter  ( i. e. 
in the Anderson width ) and eventually exhibits a maximum. The latter appears 
at values of the Anderson width which are less than the impurity  
energy level. 
Moreover, in the intermediate range of CEI this difference 
decreases with the increase of $U$.  
This dependence of $T_K$ upon the strength of CEI 
indicates that for intermediate values of $U$ the 
suppression of the charge transfer between the impurity and host
starts to be a dominant factor in the CEI influence on the singlet formation.

The paper is organized as follows: in section II the two-impurity efective 
Hamiltonian is derived, section III deals with a weak CEI
and section IV is devoted to the intermediate one. Discussions and 
Conclusions  are in section V.

\section{The effective two-impurity Hamiltonian}
\label{sec-twoimp}
We consider the Hubbard model with a magnetic impurity coupled to a lattice 
site. The Hamiltonian is
\begin{equation}
H=H_{h} + H_{imp} + H_{int}
\label{eq:H}
\end{equation}
The first term is the Hubbard Hamiltonian for the host lattice, the second 
is the impurity Hamiltonian and the third is the hybridization interaction 
between the impurity and the host. 
Figure~\ref{fig:Bethe-lattice} illustrates 
the system for the case of a Bethe lattice but the mapping discussed in this 
section is not limited to the Bethe type of lattice only.
The Hubbard Hamiltonian for the host is 
\begin{equation}
H_{h}=- \sum_{ij \sigma} (t_{ij} + \mu) c_{i \sigma}^\dagger c_{j \sigma} 
+ {U \over 2} \sum_{i, \sigma \neq \sigma'} n_{i\sigma} n_{i\sigma'}
\label{eq:H_h}
\end{equation}  
The operators $c_{i \sigma}^{\dagger}$ ( $c_{i \sigma}$) create
(annihilate) conduction electrons in spin states $\sigma$ at site
i and the corresponding density operators are
$n_{i \sigma}=c_{i \sigma}^{\dagger}c_{i  \sigma}$.
The hopping between different sites i and j is given by $t_{ij}$, $\mu$ is 
the chemical potential while $U$ is the Coulomb repulsion between two 
conduction electrons at the same site i. The impurity Hamiltonian is given 
by   
\begin{equation}
H_{imp} = \sum_{\sigma} \epsilon_f n_{f \sigma}+{U_f \over 2} 
\sum_{\sigma \neq \sigma'}n_{f \sigma} n_{f \sigma'}
\label{eq:H_imp}
\end{equation}
Here $n_{f \sigma}=f_{\sigma}^{\dagger} f_{\sigma}$ is the density 
operator for 
the impurity electron and $f_{\sigma}^{\dagger} ( f_{\sigma})$ its creation
( annihilation ) operators. The impurity level $\epsilon_f$ is taken here 
relative to the chemical potential and the Coulomb repulsion between two 
f-electrons at the impurity site is $U_f$. The hybridization interaction is 
chosen in the simple form: 
\begin{equation}
H_{int} = V \sum_{\sigma} \left( f_{\sigma}^{\dagger} c_{0 \sigma}+h.c.
\right)
\label{eq:H_int}
\end{equation}
and $0$ denotes the lattice site to which the impurity is coupled. 
We will assume a half-filled case, $<~n_{i \uparrow}~> = 
<~n_{i \downarrow}~> = 1/2$. Our treatment of the Hamiltonian of 
Eq.  (\ref{eq:H}) is based on the results of the LISA approach to the 
Hubbard model in infinite dimensions. Below we formulate some of the
 results which we use for the Anderson impurity in the Hubbard host.       

\subsection{A short overview of the LISA approach to the Hubbard model, $d 
\rightarrow \infty$}
\label{sec-LISA} 
Following \cite{LISAreview} we recall briefly the dynamical mean-field 
theory of the Hubbard model. In the limit of infinite dimensions the 
one-particle Green's function for the Hubbard Hamiltonian, Eq. 
(\ref{eq:H_h}), is local \cite{MuellerHartmann}:
\begin{equation}
G_{ij, \sigma}(i \omega_n) = \delta_{ij}G_{ii, \sigma}(i \omega_n)
\label{eq:G_h}
\end{equation}
and the usual definition 
$$G_{ij, \sigma}(i \omega_n) = 
- \int_{0}^{\beta} d\tau <Tc_{i, \sigma}(\tau)c_{j, \sigma}^{\dagger}(\tau')>
e^{i  \omega_{n} (\tau - \tau')}$$ 
 \\ 
with $\omega_n = (2n+1) \pi \beta^{-1}$ 
is used everywhere. For the paramagnetic phase which is assumed in further 
discussions and in the absence of the f-impurity we may omit indices in the
local Green's function of the Hubbard model, Eq. (\ref{eq:G_h}). The latter
 may be calculated by the means of the effective action $S_{\em{eff}}$:
\begin{eqnarray} 
S_{\em{eff}}& =& -\int_{0}^{\beta} d\tau \int_{0}^{\beta} d\tau'  
\sum_{\sigma}  c_{0, \sigma}^{\dagger}(\tau) {\cal G}_{0}^{-1}
(\tau-\tau')
c_{0, \sigma}(\tau') + \nonumber \\
& &U \int_{0}^{\beta} d\tau n_{0 \uparrow}(\tau)n_{0 \downarrow}(\tau)
\label{eq:S_eff}
\end{eqnarray}
Here Grassmann variables are used and the effective action is obtained by 
tracing out all fermions exept for a site labeled by 0 in the partition 
function: 
\begin{equation}
e^{-S_{\em{eff}}} = 
\int \prod_{i \neq 0, \sigma} Dc_{i \sigma}^{\dagger} Dc_{i \sigma}e^{-S}
\label{eq:S}  
\end{equation}
where S is the action of the Hubbard model, Eq. (\ref{eq:H_h}). The dynamic
mean-field Green's function ${\cal G}_{0}(\tau-\tau')$, 
Eq.  (\ref{eq:S_eff}), is connected to the local Green's function of the 
Hubbard model \cite{LISAreview}. The latter may be calculated by the 
self-consistent iteration procedure via the use of the definition:
\begin{equation}
G( \tau - \tau' ) = 
\frac{\int \prod_{\sigma'} Dc_{0 \sigma'}^{\dagger}
Dc_{0 \sigma'} e^{-S_{\em{eff}}} 
[- T c_{0 \sigma } ( \tau ) c_{0 \sigma }^{\dag} ( \tau')] }
{\int \prod_{\sigma'} Dc_{0 \sigma'}^{ \dag}
Dc_{0 \sigma'} e^{-S_{\em{eff}}}}   
\label{eq:G}
\end{equation}

The dynamic mean-field Green's function ${\cal G}_{0}$ may be viewed as the 
bare Green's function of the impurity electron in the auxiliary Anderson 
impurity model \cite{GeorgesKotliar,Jarrell,LISAreview}. In the latter  the 
orbital $c_{0 \sigma}^{\dagger}$ appears as the Anderson impurity and it 
posseses both the local energy level equal to $-\mu$ and the on-site Coulomb 
interaction equal to the same interaction as in the original Hubbard model, 
Eq.  (\ref{eq:H}). This fictitious Anderson impurity ( not to be confused with 
the original one in Eqs. (\ref{eq:H} -- \ref{eq:H_int})!) is coupled to a 
free-electron bath and the appropriate Hamiltonian reads 
\cite{LISAreview}:
\begin{eqnarray}                 
& &H_{AM} = \sum_{l \sigma} 
{\tilde \epsilon}_{l} a_{l \sigma}^{\dagger} a_{l \sigma}
+ \sum_{l \sigma} {\cal V}_{l} ( a_{l \sigma}^{\dagger} 
c_{0 \sigma} + h.c.) - \nonumber \\ 
& &\mu \sum_{\sigma} c_{0 \sigma}^{\dagger} c_{0 \sigma}
+ U n_{0 \uparrow} n_{0 \downarrow}
\label{eq:H_AM} 
\end{eqnarray}
The action for the Hamiltonian $H_{AM}$ is like $S_{\em{eff}}$ in Eq.  
(\ref{eq:S_eff}) with the dynamic mean-field Green's function 
${\cal G}_{0}( \tau - \tau' )$ 
given in the explicit form as a function of the parameters 
$\tilde{\epsilon}_l$ and ${\cal V}_l$. Using the Fourier transform of 
${\cal G}_{0}(\tau-\tau')$ we have \cite{LISAreview}
\begin{eqnarray}     
& &{\cal G}_{0}(i \omega_n) = \frac{\displaystyle 1}{\displaystyle i 
\omega_n + \mu - \int_{-\infty}^{\infty}
d\epsilon \frac{\displaystyle \Delta(\epsilon)}{\displaystyle i \omega_n - 
\epsilon}}; \nonumber \\[2mm]
& &\Delta(\epsilon) = \sum_{l \sigma}{\cal V}_l^{2} \delta(\epsilon - 
\tilde{\epsilon}_l)
\label{eq:G_mathcal}
\end{eqnarray}
Due to the structure of the Bethe lattice which disconnects itself into two 
separate parts by removing a lattice site, the dependence 
between ${\cal G}_{0}(i \omega_n)$ and the local Green's function for the 
original Hubbard model $G(i \omega_n)$ is simplest. It may be shown 
\cite{LISAreview} that         
\begin{equation}
G(i\omega_n) = \frac{1}{t^2} \int_{-\infty}^{\infty}
d\epsilon \frac{\Delta(\epsilon)}{i\omega_n - \epsilon} 
\label{eq:G_Bethelattice}
\end{equation}
Here 
the n.n. hopping is assumed in the Hamiltonian $H_h$, Eq. (\ref{eq:H_h}): 
$t_{ij} = t/\sqrt{d}$ for nearest neighbors and zero otherwise. Eqs. 
(\ref{eq:G_mathcal}--\ref{eq:G_Bethelattice}) together with Eq. (\ref{eq:G}) 
permit to find 
$\Delta(\epsilon)$. For the non-interacting case, 
$U = 0$, the function $\Delta(\epsilon)/t^2$ is just the density of states 
of the Bethe lattice, 
$\rho(\epsilon)$ \cite{LISAreview}:
\begin{equation} 
\rho(\epsilon) = \frac{1}{\pi t} \sqrt{1 - (\frac{\epsilon}{2t})^2}, \  
|\epsilon| \leq 2t 
\label{eq:DOS}
\end{equation}
For finite values of $U$ calculations were done in \cite{?} ( see also 
\cite{LISAreview} ) and Figure \ref{fig:DOS_U} from \cite{?} illustrates 
results of these calculations. We are interested in the paramagnetic 
phase and to 
avoid N\'{e}el order a quenched disorder may be introduced in the n. n. 
hopping. The details are presented in \cite{Rozenberg1,LISAreview} and we 
only note that by this generalisation neither the Bethe lattice 
semicircular density of states, Eq. (\ref{eq:DOS}), nor other local
 properties are changed. 

\subsection{The cavity method for the Anderson impurity in the Hubbard host}
\label{sec-cavity}
It is clear that the Anderson impurity being coupled to the site 0 destroys 
the equivalency between different sites of the Hubbard host. Let the index
$i_m, m=1,2, \ldots$ denotes one of equivalent neighbors of  the site
 0 on the Bethe lattice, m bonds away from it. Taking a site $i_m$ as the 
``cavity place'' for the dynamical mean-field treatment and tracing out all 
fermions but $c_{i_m}^{\dagger}$ the effective action 
$S_{\em{eff}}[c_{i_m}^{\dagger} c_{i_m}]$ may be calculated in close 
analogy with the pure Hubbard model. This action defines all local 
properties of the $i_m$-site and its form exactly as in Eq. (\ref{eq:S_eff}) 
with the mean-field Green's function ${\cal G}_{0}^{(i_m)}(\tau - \tau')$ 
instead of the ${\cal G}_{0}(\tau - \tau')$. 
By straightforward calculations \cite{Benny} it may be shown that
\begin{eqnarray}
{\cal G}_{0}^{(i_1)}(i \omega_n)& =& i \omega_n + \mu + 
\sum_{\stackrel{i ~ n.n. ~ of ~ i_{1}}{i \neq 0}} 
\frac{t^{2}}{d} G_{i i } (i \omega_n)  + \nonumber \\[2mm]
& & \frac{t^{2}}{d} G_{0 0}(i \omega_n ) \nonumber \\[2mm]
{\cal G}_{0}^{(i_m)}(i \omega_n&) =& i \omega_n + \mu + 
\sum_{ i ~ n.n. ~ of ~ i_{m}} \frac{t^{2}}{d} G_{i i } (i \omega_n); m \neq 1
\label{eq:G_mathcal_i_m}  
\end{eqnarray}
It is obvious that for the case of infinite dimensions Eqs. 
(\ref{eq:G_mathcal_i_m}) are solved by the ${\cal G}_{0}(i \omega_n)$ and 
$G(i \omega_n)$ , Eqs. (\ref{eq:G_mathcal}, \ref{eq:G_Bethelattice}) of the
pure Hubbard model. In fact an impurity can not break the equivalncy between
different sites $i \neq 0$ of the Hubbard host and does not influence their 
local properties.

To calculate the correlation function of the impurity electron we trace out
all fermions exept two, $c_{0 \sigma}^{\dagger}$ 
and $f_{\sigma}^{\dagger}$, in the partition function and define the effective 
action $S_{\em{eff}}[0,f]$:
\begin{eqnarray}
& &e^{-S_{\em{eff}}[0,f]} = \int \prod_{i \neq 0, \sigma}
Dc_{i \sigma}^{\dagger}
Dc_{i \sigma} e^{S} \nonumber \\
& &S = \int_{0}^{\beta} d\tau (\sum_{i \sigma}c_{i \sigma}^{\dagger}(\tau) 
\partial_{\tau} c_{i \sigma}(\tau) + \sum_{\sigma} f_{ \sigma}^{\dagger}
(\tau) \partial_{\tau} f_{ \sigma}(\tau) + \nonumber \\
& &H_h(\tau) + H_{imp}(\tau) + H_{int}(\tau))
\label{eq:S_eff[0,f]_define}  
\end{eqnarray}
Here $H_h$, $H_{imp}$, $H_{int}$ are from Eqs. (\ref{eq:H_h}), 
(\ref{eq:H_imp}), (\ref{eq:H_int}) respectively. Repeating the same 
calculations which lead to $S_{\em{eff}}$ of Eq. (\ref{eq:S_eff}) 
 for the Hubbard model without impurity \cite{LISAreview}, we obtain for the 
Bethe lattice of infinite connectivity
\begin{eqnarray}
& &S_{\em{eff}}[0,f] = -\int_{0}^{\beta} \! d\tau \! \int_{0}^{\beta} \! 
d\tau' \sum_{\sigma}
c_{0, \sigma}^{\dagger}(\tau) {\cal G}_{0}^{-1}(\tau-\tau')
c_{0,\sigma}(\tau')  \nonumber \\
& &+ \int_{0}^{\beta} d\tau ( Un_{0 \uparrow}(\tau)n_{0 \downarrow}(\tau) + 
\sum_{\sigma}
f_{ \sigma}^{\dagger}(\tau) \partial_{\tau} f_{ \sigma}(\tau) + \nonumber \\
& &H_{imp}(\tau) + H_{int}(\tau)) 
\label{eq:S_eff[0,f]}  
\end{eqnarray}
Note that here the dynamic mean-field Green's function 
${\cal G}_{0}(\tau-\tau')$ coincides with the Green's function for the
 pure Hubbard model, Eq. (\ref{eq:G_mathcal}).

To study the Kondo effect it is convenient to work with the effective 
Hamiltonian, $H_{\em{eff}}[0,f]$, which produces the same action as in 
Eq. (\ref{eq:S_eff[0,f]}). 
Because the Hubbard host is represented in Eq. (\ref{eq:S_eff[0,f]}) exactly
like in the case of the Hubbard model without impurity it is easy to see 
that 
$$H_{\em{eff}}[0,f] = H_{AM} + H_{imp} + H_{int}$$.
Using Eqs. (\ref{eq:H_imp}), (\ref{eq:H_int}), (\ref{eq:H_AM}) we obtain 
finally
\begin{eqnarray}      
H_{\em{eff}}[0,f]& =& \sum_{l \sigma} 
\tilde{\epsilon}_{l} a_{l \sigma}^{\dagger} a_{l \sigma}
+ \sum_{l \sigma} {\cal V}_{l} ( a_{l \sigma}^{\dagger} 
c_{0 \sigma} + h.c.) -  \nonumber \\ 
& & \mu \sum_{\sigma} c_{0 \sigma}^{\dagger} c_{0 \sigma}
+ Un_{0 \uparrow}n_{0 \downarrow} + \nonumber \\                         
& & \sum_{\sigma} \epsilon_f n_{f \sigma}+{U_f \over 2} 
\sum_{\sigma \neq \sigma'}n_{f \sigma} n_{f \sigma'} + \nonumber \\
& &V \sum_{\sigma} ( f_{\sigma}^{\dagger} c_{0 \sigma}+h.c. )     
\label{eq:H_eff[0,f]}    
\end{eqnarray}
The above Hamiltonian describes the two-impurity system in which 
the first one, $c_{0 \sigma}^{\dagger}$, represents, together with the 
free-electron bath, the Hubbard host ( see subsection~
\ref{sec-LISA} ) and the second impurity is coupled to the first one only.
It is still a non-trivial model but because 
the influence of the impurity on the fitting set of parameters 
$\tilde{\epsilon}_{l}$ and ${\cal V}_{l}$ is negligible we may use the
function $\Delta(\epsilon)$, Eq. (\ref{eq:G_mathcal}), as it is calculated  
for the Hubbard model. The function $\Delta(\epsilon)$ 
depends of course on the type of the lattice.  
In the following Section we consider the Kondo effect as it emerges from
the Hamiltonian, Eq. (\ref{eq:H_eff[0,f]}), for the case of the Bethe 
lattice of infinite connectivity with the function $\Delta(\epsilon)$ taken 
from \cite{?}. We assume that the energy $U_f$ is much larger than other 
energetical scales in the problem and will take it infinitely large to exclude 
the double occupation of the impurity site.

\section{Weak Correlated Host}
\label{sec-weak}
To take into account the absence of double f-occupation we
follow \cite{Coleman} and introduce the slave boson field $b^{\dagger}, b$
with the restriction
\begin{equation}
b^{\dagger}b + \sum_{\sigma} c_{0 \sigma}^{\dagger} c_{0 \sigma} = 1
\label{eq:rstrctn}
\end{equation}
The two-impurity Hamiltonian has to be modified in the well known fashion
\cite{Coleman,Bickers} and its form now is  
\begin{eqnarray}       
& &H_{\em{eff}}[0,f,b] = \sum_{l \sigma} 
\tilde{\epsilon}_{l} a_{l \sigma}^{\dagger} a_{l \sigma}
+ \sum_{l \sigma} {\cal V}_{l} ( a_{l \sigma}^{\dagger} 
c_{0 \sigma} + h.c.) - \nonumber \\ 
& &\mu \sum_{\sigma} c_{0 \sigma}^{\dagger} c_{0 \sigma}
+ Un_{0 \uparrow}n_{0 \downarrow} + \nonumber \\                         
& &\sum_{\sigma} \epsilon_f n_{f \sigma} + \lambda b^{\dagger}b
+ V \sum_{\sigma} ( f_{\sigma}^{\dagger} b c_{0 \sigma}+h.c.)
\label{eq:H_eff[0,f,b]}    
\end{eqnarray}
Here $\lambda$ is the slave boson chemical potential which in the final stage
of calculations has to be put to $-\infty$ in order to satisfy the 
restriction of Eq. (\ref{eq:rstrctn}). 

We begin with free electrons in the original Hamiltonian, 
Eq. (\ref{eq:H}), that is $U = 0$. Using the NCA in the lowest $1/N$ 
approximation the boson and f-pseudofermion self-energies may be calculated 
by the use of the Hamiltonian $H_{\em{eff}}[0,f,b]$ and compared with the 
well known results \cite{Coleman,Bickers}. Figures \ref{fig:bubbles}  a, b 
depict these self-energies. The propagator of the first impurity electron in 
our two-impurity 
system ( states $c_{0\sigma}^{\dagger}$ ) is labeled by $G_{AM}(i \omega_n)$. 
For
$U = 0$ it is easy to calculate that
\begin{equation}
G_{AM}(i \omega_n) = 
\frac{1}{\displaystyle i \omega_n + \mu -\sum_{l}
\frac{\displaystyle {\cal V}_{l}^{2}}{\displaystyle i \omega_n - 
\tilde{\epsilon_l}}}
\label{eq:G_AM_U=0}
\end{equation}       
The self-consistent mapping of the original system with $U$ = 0 onto the 
two-impurity one means that the local Green's function $G(i \omega_n)$ 
coincides with $G_{AM}(i \omega_n)$. So we obtain 
\begin{equation}
G_{AM}(i\omega_n) \equiv G(i\omega_n) = \int_{-\infty}^{\infty}  
d\epsilon \frac{\rho(\epsilon)}{i \omega_n - \epsilon + \mu}
\label{eq:G_AM=G}
\end{equation}
Here $\rho(\epsilon)$ is the DOS of the host lattice, Eq. (\ref{eq:H_h}).  
Substituting in the expressions for the self-energies of Figures 
\ref{fig:bubbles} a, b instead of the propagator $G_{AM}(i \omega_n)$ 
the Green's function $G(i \omega_n)$ from 
Eq. (\ref{eq:G_AM=G}) one can recover, as it is expected for the 
non-interacting case, self-energy expressions as they 
emerge by the direct application of the NCA to the original Hamiltonian 
Eq. (\ref{eq:H}) \cite{Coleman,Bickers}. The Kondo temperature
$T_K$ changes exponentially as a function of the coupling and in the lowest 
iteration of the NCA it has the form \cite{VarmaYafet}
\begin{equation}
T_K = 2t \exp{\frac{\displaystyle \pi \epsilon_f}{\displaystyle 2 \Gamma}}
\label{eq:T_K}   
\end{equation}
Here  $\Gamma = \pi V^2 \rho(0)$ is the Anderson width.

The interaction case, $U \neq 0$,  may also be considered in the frame of the
NCA. In the lowest $1/N$ approximation  the slave boson and 
f-pseudofermion self-energies acquire  
vertex corrections which take into account the CEI. These vertex corrections
$\Gamma^{U}(1,2;3,4)$
were discussed in \cite{TZZ2} and shown in Figures \ref{fig:bubbles} c, d.
Now neither the propagator $G_{AM}$ has the simple form of Eq. 
(\ref{eq:G_AM_U=0}) nor $G$ is of the non-interacting nature of Eq. 
(\ref{eq:G_AM=G}). 
However, for weak correlations the self-energies, 
Figures \ref{fig:bubbles} c, d,  may be calculated \cite{TZZ2}. In this case
of small U one may neglect the mass renormalization of band electrons 
\cite{KhaliullinFulde} and use for $G_{AM}$ Eq. (\ref{eq:G_AM=G}), besides 
$\Gamma^{U}(1,2;3,4) = U$. With these approximations the 
self-energies of  Figures \ref{fig:bubbles}  c, d are exactly the same as 
calculated in \cite{TZZ2} for the Hamiltonian, Eq. (\ref{eq:H}), in the usual 
three-dimensional case in the {\em local} approximation. 
According to \cite{TZZ2} Eq. (\ref{eq:T_K}) is
preserved ( in the lowest iteration of the NCA ) but the parameters 
$\epsilon_f$ and $\Gamma$ undergo the renormalization ( both $|\epsilon_f|$ 
and $\Gamma$ increase linearly with $U$ ) leading to the enhancement 
of $T_K$. Details of this renormalization may be found in \cite{TZZ2}.
Note that in this section we did not use any specific properties of the Bethe
lattice and its results are valid for any host lattice.  
It would be interesting to go in the NCA beyond the linear in the Hubbard 
$U$ approximation but here we will apply the variational 
principle to handle the effective two-impurity Hamiltonian for intermedium 
values of $U$.
\section{Intermediate Interactions: Variational Approach}
\label{sec-variational} 

By the intermediate interaction case we mean the range of the ratio $U/2t$ 
for which the host is still in the metallic regime, i.e. much before the
metal-insulator transition takes place. From Figure \ref{fig:DOS_U} for the 
Bethe lattice which is considered throughout this section we take 
$U/2t \stackrel {<}{\sim} 3$ somewhat loosely for the
upper limit . The lower limit in our case is 
dictated by the variational functions implemented here which put this limit
at $U/2t \sim 1$ ( see subsection \ref{sec-numerics} ).

We construct here variational functions for the singlet and triplet states
and find the ground singlet and lowest triplet energies. Essentially, we
 follow ideas of $1/N$ variational treatment of the Anderson impurity 
Hamiltonian developed in \cite{GS1,GS2}. It is convenient to introduce a new
representation for the bath states in Eq. (\ref{eq:H_eff[0,f]}):
\begin{eqnarray}
\Psi_{\epsilon \sigma}^{\dagger}& =&{\cal V}^{-1}(\epsilon) \sum_{l \sigma}
{\cal V}_{l}^{2} \delta( \epsilon - \tilde{\epsilon_l} ) a_{l \sigma}^{\dag}
\nonumber \\ 
{\cal V}(\epsilon)& \equiv & \sqrt{\Delta(\epsilon)}
\label{eq:Psi}   
\end{eqnarray} 
Here $\Delta(\epsilon)$ is from Eq. (\ref{eq:G_mathcal}). Using this definition
 the Hamiltonian $H_{\em{eff}}[0,f]$ may be rewritten as follows:  
\begin{eqnarray}
H_{\em{eff}}[0,f]& =& \sum_{\sigma} \int \; d\epsilon ( \epsilon 
\Psi_{\epsilon \sigma}^{\dagger} \Psi_{\epsilon \sigma} + {\cal V}(\epsilon)
( \Psi_{\epsilon \sigma}^{\dagger} c_{0 \sigma} + h.c. )) - \nonumber \\ 
& &\sum_{\sigma}( \mu n_{0,\sigma}- \epsilon_f n_{f \sigma})  + 
V \sum_{\sigma} ( f_{\sigma}^{\dagger} c_{0 \sigma}+h.c. ) + \nonumber \\ 
& &Un_{0 \uparrow}n_{0 \downarrow} +{U_f \over 2} 
\sum_{\sigma \neq \sigma'}n_{f \sigma} n_{f \sigma'} 
\label{eq:H_Psi}    
\end{eqnarray}           
Note that since $\mu = U/2$ the $c_0$-impurity is described by 
the symmetric Anderson model while the double occupancy of the f-level is 
forbidden ( $U_f \rightarrow \infty$ ).

\subsection{Variational functions}
\label{sec-functions}
Figure \ref{fig:boxes_s} schematically represents the singlet state 
variational function. We have two groups of states: one contains four states 
in which the f-level is unoccupied, the other is of six states in which 
the f-level is singly occupied. For vanishing hybridization coupling $V = 0$ 
two groups are 
decoupled and the first one is just the variational function of the symmetric 
Anderson impurity. We want to keep only the lowest contributions in $1/N$  
which come from the states without electron-hole pairs \cite{GS1,GS2}. 
For the symmetrical Anderson impurity the occupation of the empty  state 
has to equal that of the doubly-occupied one which  
may be achieved only by including states with one electron-hole pair at least 
\cite{GS2}. We included only one such state 
( $\phi_3$ in Figure \ref{fig:boxes_s} ) and we checked that the four-state 
variational function for the symmetric Anderson impurity produces the required 
equality between occupancies of the empty and doubly-occupied states. 
There are two additional states with one electron-hole pair \cite{GS2}. 
Including them, however,would make calculations too cumbersome.
Being interested in the qualitative influence of the Hubbard $U$ on the
Kondo effect we limited ourself to the lowest possible combination of states 
which is expected to give correct qualitative results at least. The second 
group of states ( states 4 - 9 in Figure \ref{fig:boxes_s} ) are choosen from
the same considerations. 

We denote states by $\phi_i, \; i = 0, 1, \ldots ,9$, so $\phi_0 = |0>$ 
represents the vacuum ( full Fermi sea and empty local states ), 
$\phi_1(\epsilon) = 1/\sqrt{2} \sum_{\sigma} \Psi_{\epsilon, \sigma}^{\dag} 
c_{0 \sigma}^{\dag}|0>,
 \ldots ,\phi_9(\epsilon_1, \epsilon_2, E) = 1/\sqrt{2} \sum_{\sigma} 
\Psi_{\epsilon_1, \sigma}^{\dag}\Psi_{\epsilon_2, \sigma}^{\dag} 
\Psi_{E, \sigma}^{\dag} f_{\sigma}^{\dag}|0>$. All $\phi$-functions  are 
listed in the Appendix \ref{sec-tables}. In this basis the singlet ground 
state of the Hamiltonian Eq. (\ref{eq:H_Psi}) has the following form:
\begin{eqnarray}
\lefteqn{ \psi_{\em{s}} = {\cal N}[ \; \phi_0 + \int_{-D}^{0} d\epsilon_1 (} 
\nonumber \\
& &r_1(\epsilon_1) \phi_1 + \int_{-D}^{\epsilon_1} d\epsilon_2 r_2(\epsilon_1
, \epsilon_2)\phi_2 + \int_{0}^{D} dE r_3(\epsilon_1, E)\phi_3 + \nonumber \\
& &r_4(\epsilon_1) \phi_4 + \int_{-D}^{0}\; d\epsilon_2 r_5(\epsilon_1
, \epsilon_2)\phi_5 + \int_{0}^{\epsilon_1} d\epsilon_2 r_6(\epsilon_1,
\epsilon_2)\phi_6 + \nonumber \\
& &\int_{-D}^{\epsilon_1} \! \! \! d\epsilon_2 \! \! \int_{D}^{0}\! \! \! 
d\epsilon_3 r_7(
\epsilon_1, \epsilon_2, \epsilon_3) \phi_7  + \int_{-D}^{0}\! \! \! 
d\epsilon_2
\!  \! \int_{0}^{D}\! \! \! dE r_8(\epsilon_1, \epsilon_2, E) \phi_8 
\nonumber \\ 
& & +\int_{-D}^{\epsilon_1} \! \! \! d\epsilon_2 
\! \! \int_{0}^{D} \! \! \! dE r_9(\epsilon_1, \epsilon_2, E) \phi_9 \;) \; ]
\label{eq:singlet}
\end{eqnarray}
Here ${\cal N}$ is the normalization factor and $r_i$ are superposition 
functions which have to be determined by the energy minimization. For 
convenience, the energies of free electrons are denoted here by $\epsilon$ 
for a 
negative part of their spectrum and by $E$ for the positive part and $D 
\equiv 2t$. Because the replacement $\epsilon_1 \leftrightarrow \epsilon_2$
for $\phi_2(\epsilon_1, \epsilon_2)$, $\phi_6(
\epsilon_1, \epsilon_2)$ and $\phi_9(\epsilon_1, \epsilon_2, E)$ does not 
give new 
independent functions the upper limit of appropriate inner integrals is 
equal to $\epsilon_1$. The first four out of ten 
integral equations connecting different $r_i, i = 0,1 \ldots ,9$ are
\begin{equation}
r_0 = \frac{\int_{-D}^{0} d\epsilon r_1(\epsilon)}{\Delta E_{\em{s}}} 
\label{eq:intgrleq1}
\end{equation}      
and
\begin{eqnarray}
r_1(\epsilon)& =& [ \; V r_2(\epsilon) +\sqrt{2} {\cal V}(\epsilon)r_0 + 
\int_{-D}^{0} \! \! \! \ d\epsilon_1{\cal V}(\epsilon_1)\tilde{r}_2(\epsilon,
\epsilon_1)+ \nonumber \\
& &\int_{0}^{D}\! \! \! dE {\cal V}(E) r_2(\epsilon, E) \; ]/(\; \Delta 
E_{\em{s}} + \epsilon + \mu \; ) \nonumber \\
\tilde{r}_2(\epsilon, \epsilon_1) &=& [\;{\cal V}(\epsilon_1)r_1(\epsilon_2) 
+ {\cal V}(\epsilon_2)r_1(\epsilon_1) \! + \! \nonumber \\ 
& &V (\; r_5(\epsilon_1, \epsilon_2) + 
r_5(\epsilon_2, \epsilon_1)\;) \;]/( \;\Delta E_{\em{s}} + \epsilon_1 + 
\epsilon_2 \;) \nonumber \\
r_3(\epsilon, E) &=& {\cal V}(E)r_1(\epsilon)/(\;\Delta E_{\em{s}} + 
\epsilon - E \; ) 
\label{eq:intgrleq2}
\end{eqnarray}      
while the rest of the integral equations may be found in Appendix 
\ref{sec-intgrleq}. Here $\Delta E_{\em{s}}$ is the energy difference 
between the singlet ground state energy and the energy of the filled Fermi 
sea of the effective conduction band, Eq.(\ref{eq:H_Psi}); 
$\tilde{r}_2(\epsilon, 
\epsilon_1) =  r_2(\epsilon, \epsilon_1)$ if $\epsilon \geq \epsilon_1$ and
$\tilde{r}_2(\epsilon, \epsilon_1) =  r_2(\epsilon_1, \epsilon)$ if 
$\epsilon 
\leq \epsilon_1$. Before we procceed with the numeric variational 
calculations let us discuss the variational functions for magnetic states.

We consider triplet, $S = 1$, states. Because of the rotational symmetry it 
is sufficient for our purposes to deal with the state of $S_z = 1$. The basis 
of this state may be generated from the singlet basis of Appendix 
\ref{sec-tables} by applying to them the operator 
$\Psi_{\epsilon \downarrow} 
 c_{0 \uparrow}^{\dag}$. The magnetic basis consists of functions $\Phi_i, i =
 1, \ldots ,12$ and the first three are $\Phi_1(\epsilon) = \Psi_{\epsilon 
\downarrow} c_{0 \uparrow}^{\dag} |0>$, $\Phi_2(\epsilon_1, \epsilon_2) = 
\Psi_{\epsilon_1 \downarrow} \Psi_{\epsilon_2 \downarrow} 
c_{0 \downarrow}^{\dag} c_{0 \uparrow}^{\dagger} |0>$ and 
$\Phi_3(\epsilon, E) =\Psi_{\epsilon \downarrow}
\Psi_{E \uparrow}^{\dagger}|0~>$. These three states compose the $S_z = 1$ 
state of the symmetric Anderson impurity when it is decoupled from the rest 
of the system,  Eq. (\ref{eq:H_Psi}). Other basis states are written in 
Appendix \ref{sec-tables}. The triplet $S_z = 1$ variational function therefore
reads:
\begin{eqnarray}
\lefteqn{\psi_{\em{t}}= {\cal N}[\int_{-D}^{0} d\epsilon_1 ( } \nonumber \\
& &R_1(\epsilon_1)\Phi_1 \! + \! \int_{-D}^{\epsilon_1} \! \! \! d\epsilon_2 
R_2(\epsilon_1,\epsilon_2)\Phi_2 \!\! + \! \! \int_{0}^{D} \! \! \! dE 
R_3(\epsilon_1,E)\Phi_3 + \nonumber \\
& &R_4(\epsilon_1)\Phi_4 \!\! + \!\! \int_{-D}^{\epsilon_1} \! \! \!
d\epsilon_2
R_5(\epsilon_1,\epsilon_2)\Phi_5 \! + \! \int_{-D}^{0} \! \! \! d\epsilon_2
R_6(\epsilon_1,\epsilon_2)\Phi_6  + \nonumber \\
& &\int_{-D}^{\epsilon_1}\! \! \!d\epsilon_2R_7(\epsilon_1,\epsilon_2)\Phi_7 
\! \!+\! \! \int_{-D}^{0}\! \! \!d\epsilon_2\! \! \!\int_{-D}^{\epsilon_1}
\! \! \!d\epsilon_3R_8(\epsilon_1,\epsilon_2, \epsilon_3)\Phi_8  \; \; \; + 
\nonumber \\ 
& &\int_{-D}^{\epsilon_1}\! \! \!d\epsilon_2\! \! \!\int_{-D}^{\epsilon_2}
\! \! \!d\epsilon_3R_9(\epsilon_1,\epsilon_2, \epsilon_3)\Phi_9 \;+ 
\nonumber\\
& &\int_{-D}^{\epsilon_1}\! \! \!d\epsilon_2\! \! \!\int_{0}^{D}
\! \! \!dER_{10}(\epsilon_1,\epsilon_2, E)\Phi_{10}   + 
\nonumber \\
& &\int_{-D}^{0}\! \! \!d\epsilon_2\! \! \!\int_{0}^{D}
\! \! \!dER_{11}(\epsilon_1,\epsilon_2, E)\Phi_{11} \nonumber + \\
& &\int_{-D}^{\epsilon_1}\! \! \!d\epsilon_2\! \! \!\int_{0}^{D}
\! \! \!dER_{12}(\epsilon_1,\epsilon_2, E)\Phi_{12}\; )\;]
\label{eq:psi-triplet}
\end{eqnarray}
Here upper limits of some integrals are choosen to avoid double counting of 
states and functions $R_i$ must be obtained from the minimization of the 
triplet energy. As it is seen from the list of the functions in Appendix 
\ref{sec-tables} the $S_z =1$ basis is composed from three groups of states. 
The first group contains the above written symmetric Anderson impurity 
states, in the two others the real impurity level is occupied, the difference 
between the second and the third group being the origin of the triplet. In 
the second the triplet originates from the real impurity and a conduction
electron and in the third group it is from the symmetric Anderson impurity 
and a conduction electron. Integral equations which connect different 
functions $R_i$ are collected in Appendix \ref{sec-intgrleq}. In the next 
subsection we suggest an effective numerical iteration method which was 
implemented here for carring out the energy minimization both for the 
singlet and for the triplet.
\subsection{An iteration method for the energy minimization and results}
\label{sec-numerics} 
The direct numerical diagonalization of the two systems of linear integral 
equations, Eqs. (\ref{eq:intgrleq1}), (\ref{eq:intgrleq2}), 
(\ref{eq:ve} - \ref{eq:vh2} ) for the singlet state and (\ref{eq:vam} - 
\ref{eq:vh3m} ) for the magnetic state, would be a 
very time consuming task. To find the energy shift 
$\Delta E_{\em{s}}$ together with 
the $r_i$ functions we solve  the integral equations system of Eqs.
(\ref{eq:intgrleq1}), (\ref{eq:intgrleq2}), (\ref{eq:ve} - \ref{eq:vh2}) 
numerically for an arbitrary given value of $\Delta E_{\em{s}}$, say 
$\Delta E$ . This is done by the controlled iteration procedure which 
starts with a most resonable guess about the initial $r_i$ functions. 
A fast convergence is usually obtained by choosing all but two initial $r_i$  
equal to zero and by selecting these two from states which are
 expected to contribute significally to the superposition $\psi_{\em{s}}$. 
Let $r_i^{\Delta E}$ are functions calculated in the above procedure. Then 
we obtain by the use of Eq. (\ref{eq:intgrleq1}) the following function 
$f(\Delta E)$: 
\begin{equation}
f(\Delta E) \equiv \frac{\int_{-D}^{0} d\epsilon r_1^{\Delta E}(\epsilon)}
{r_0^{\Delta E}} 
\label{eq:fixedpoint_f}
\end{equation}      
In view of the arbitrariness of $\Delta E$, $f(\Delta E)$ does not coincide 
with $\Delta E$, the equality between them is realized by 
$\Delta E_{\em{s}}$ ( see Eq. (\ref{eq:intgrleq1})). 
So the latter may be seen as a 
minimal possible fixed point of the function $f$, Eq. (\ref{eq:fixedpoint_f}):
$$f(\Delta E_{\em{s}}) = \Delta E_{\em{s}}.$$ 
This fixed point may be found after a few trials. Note that this 
scheme does not require the normalization 
of the $\psi_{\em{s}}$, Eq.(\ref{eq:singlet}), to be fulfilled
in each stage of iterations. To find the energy shift 
$\Delta E_{\em{t}}$ between the lowest magnetic state and the energy of the 
filled Fermi sea of the conduction band of Eq. (\ref{eq:H_Psi}) the 
analogical iteration method is applied to the above magnetic state, 
Eq. (\ref{eq:psi-triplet}). An illustration of the iteration procedure and 
some details of it are given in Appendix \ref{sec-iterations}. 

It is clear from general considerations that values of $U$ here have to be 
limited from below just because of the restricted basis which is used ( see 
Figure \ref{fig:boxes_s} ). In addition, in our case we are interested in the 
Kondo limit of almost filled states $\phi_5$ and $\phi_6$. Therefore 
the local energy $U/2$ of the Anderson impurity 
with states $c_{0,\sigma}^{\dag}$ in the two-impurity Hamiltonian, Eq. 
( \ref{eq:H_Psi} ), has to be much larger than the maximal coupling parameter
$\Gamma_1$ = $\pi {\cal V}^2(0)$. By the definition of 
${\cal V}(\epsilon)$, Eq. (\ref{eq:Psi}), this gives :
$$U/D \gg 1/4$$
The hybridization coupling parameter for the real impurity in the 
two-impurity
Hamiltonian, Eq. ( \ref{eq:H_Psi} ), may be defined in the usual way from the
original Hamiltonian, Eq. ( \ref{eq:H} ), as the Anderson width 
$\Gamma$ =$\pi V^2 \rho(0)$ with $\rho(0)$ = $2/\pi D$ ( see Figure 
\ref{fig:DOS_U} ). The Kondo limit for the real impurity is given then by 
the condition:
$$|\epsilon_f| > 2V^2/D.$$
Two values of $\epsilon_f/D$ = - 0.67 and - 0.3 were used in our 
calculations. Figure \ref{fig:curves_U=1} presents the dependence of the 
energy difference $T_K = \Delta E_{\em{t}} - \Delta E_{\em{s}}$ on $V^2$ 
for both values of $\epsilon_f$ and $U/D$ = 1 as calculated by the described 
above iteration method. 

Figure \ref{fig:curves_U=2} compares $T_K$ as a function of $V^2$ 
for different values of $U$ and $\epsilon_f/D$ = - 0.3.
Figure \ref{fig:curve_U} 
illustrates the decrease of $T_K$ with the U increase. Here for the 
comparison the standard Kondo temperature from Eq. (\ref{eq:T_K}) is also 
shown. 
\section{Discussions and Conclusions}
\label{sec-DC}
Summarizing the above we note  that i) weak coupling between conduction 
electrons does not destroy the usual exponential Kondo scale ( for 
$\Gamma \ll \epsilon_f$ ) but renormalizes the parameters $\Gamma$ and 
$\epsilon_f$; ii) the exponential scale of $T_K$ is lost already for $U/D$ = 1.
Moreover, for small $\Gamma$ the Kondo temperature $T_K$ is 
proportional to $V^2$ ( see Figures \ref{fig:curves_U=1} and  
\ref{fig:curves_U=2} ); iii) there is a pronounced maximum on Figure 
\ref{fig:curves_U=2} for $U$ = 2, 2.5 and an indication on a broad one in 
Figure \ref{fig:curves_U=1} for $U$ = 1 and iv)in the range of $U \geq D$ the
$T_K$ decreases with the increase of $U$. 

The enhancement of $T_K$ in the Kondo regime by a weak coupling between 
conduction electrons is understood ( see \cite {KhaliullinFulde,TZZ2} ) and 
is caused by the reduced probability of finding doubly occupied and empty 
lattice sites in the correlated system. This leads to the increased number of
uncompensated conduction electron spins and to increase of the effective 
hybridisation. Our two-impurity model for $d \rightarrow \infty$ gives just the
 same result. 

The loss of the exponential scale for intermediate $U/D$ = 2 and 2.5 does
not come as a surprise because for these values the DOS of Figure 
\ref{fig:DOS_U} has the typical three peak structure of the Anderson impurity 
and the Hubbard host excitations in this case are not of the Fermi-liquid 
character \cite{LISAreview}. However, for $U$ = 1 there are 
no gaps in the DOS and $\epsilon_f$ lies within the metallic looking DOS.
Nevertheless also this case is qualitatively different from the 
non-interacting one. This
difference can be seen in the frame of our two-impurity model. 
Indeed when the real impurity is decoupled, $V$ = 0, the rest of the system 
is just 
the symmetric Anderson impurity embedded in the bath. At $T = 0$ and provided 
that $U \gg \Gamma_1$ this system  posseses the 
Kondo temperature which may be viewed as the energy difference between the 
ground singlet and lowest relevant magnetic states \cite{footnote4} of the 
symmetric Anderson model. This Kondo temperature for the symmetric Anderson 
impurity is refereed to as $T_{K}^{AM}$ and it was calculated in Appendix 
\ref{sec-iterations} for several values of $U$ ( see Appendix  
\ref{sec-iterations} ). 
By coupling the real Anderson 
impurity to the system with a precursory non-zero energy difference 
between the ground singlet and the lowest relevant magnetic states we may 
expect for small coupling $2V^2/D \ll T_{K}^{AM}$ a perturbative linear 
dependence between $T_K$ and $V^2$. Using values of $T_{K}^{AM}$ from 
Appendix \ref{sec-iterations}  we obtain regions of 
linear dependence which are in a fair agreement with Figures 
\ref{fig:curves_U=1} and 
\ref{fig:curves_U=2}. For small enough $U < \Gamma_1$ the symmetric Anderson 
impurity enters in the non-magnetic regime of the resonance level ( see 
\cite{TsvelickWiegmann} ). Using $\Gamma_1 = \pi {\cal V}^2(0)$ and Eqs. 
(\ref{eq:DOS}) and (\ref{eq:Psi}) we see that this happens at $U/D \sim 1/2$ 
and we expect that the usual Fermi-liquid picture of the Kondo effect emerges
below this value of $U/D$. 
Unfortunately, as was discussed in the subsection \ref{sec-numerics}, we 
cannot  treat this region of smaller $U$ in the frame of the above variational 
approach. So the physics of the crossover from the Fermi-liquid regime with 
the exponential Kondo scale to the new regime where the Kondo exponential 
scale is lost is waiting for further investigations, perhaps in the frame of 
the diagrammatic approach but beyond the weak coupling which was discussed in  
Section \ref{sec-weak}.  

In general a non-monotonic behaviour of $T_K$ as a function of $V^2$ 
may be expected at $\Gamma \stackrel {<} {\sim} |\epsilon_f|$ when the 
transition from the Kondo regime to the mixed valence regime 
occurs \cite{footnote5}. The maximum of the upper curve with $|\epsilon_f| = 
0.3$, Figure \ref{fig:curves_U=1},  occurs at $\Gamma$ = $2V^2/D 
\sim 0.2$, and is in accordance with this expectation. However the lower 
curve of this figure which corresponds to the same $U/D = 1$ and about 
twice larger $|\epsilon_f|$ seems to saturate in the same region as the 
upper curve.
In view of  the specific form of the DOS of Figure
\ref{fig:DOS_U} one has to be cautious to directly use the criteria taken 
from the non-interacting case. It is especially true for the cases of larger 
$U$ =2, 2.5 and $\epsilon_f$ = -0.3 on Figure \ref{fig:curves_U=2}. The 
maximum appears there at $\Gamma \sim 0.1$ and 0.04 correspondingly showing a 
strong dependence of $T_K$ upon $U$. In both last cases the local bare 
level $\epsilon_f$ is on the edge of the central portion of DOS, 
Figure \ref{fig:DOS_U}, which narrows much with $U$. This narrowing 
of DOS determines the $U$-dependence of $T_K$.

The decrease of $T_K$ as $U$ increases is expected for sufficiently large $U$
because the charge transfer from the impurity to the lattice eventually 
will be inhibited by the energy cost of the double occupancy of the lattice 
sites. The most interesting region of $U$ values, where the cross-over from 
the exponential Kondo scale to the perturbative one takes place, has yet to be 
explored. It would be desirable to study the entire range of $U$ by one 
method of 
treatment. It would be interesting also to look into the experimental 
aspects of the Kondo effect as a tool for detecting non-Fermi-liquid  
properties of a strongly correlated metal. 

In conclusion, we reduced the treatment 
of the Kondo effect in the Hubbard host of infinite dimensions to study of 
a simpler two-impurity Hamiltonian. The weakly correlated case was treated 
by the NCA and relations to previous studies of the Anderson impurity in a 
correlated host in 2d and 3d were shown. The variational treatment of the 
two-impurity Hamiltonian for intermediate interactions reveals a 
non-exponential behaviour of the Kondo scale and qualitative explanations 
for this behavior were proposed. 
\section{Acknowledgement}
This work was supported by the Israel Science Foundation administered  by 
the Israel Academy of Sciences and Humanities. V. Z. is grateful to the
Max-Planck-Institute PKS for hospitality. The authors would like to thank 
G. Kotliar for education in the LISA during his lectures in Jerusalem and
M. Rozenberg for sending to us his numeric data of DOS of the Hubbard model 
on the Bethe lattice. Useful discussions with G. Khaliullin and G. Zwicknagl 
are acknowledged.
\appendix{Tables of functions}
\label{sec-tables}
The basis of the singlet ground state, Eq. (\ref{eq:singlet}), follows:
\begin{eqnarray}
\phi_{0} &=& |0> \label{eq:m} \\
\phi_{1}( \epsilon ) &=& \frac{1}{\sqrt{2}} \sum_{\sigma}
\Psi_{\epsilon \sigma} c_{0 \sigma}^{\dag} |0> 
\label{eq:A} \\
\phi_{2} ( \epsilon_{1} , \epsilon_{2} )  &=& \sqt \sum_{\sigma}
\Psi_{\epsilon_{1} \sigma} \Psi_{\epsilon_{2} - \sigma}
c_{0 \sigma}^{\dag} c_{0 - \sigma}^{\dag} |0> 
\label{eq:B} \\
\phi_{3} ( \epsilon , E )  &=& \sqt \sum_{\sigma} \Psi_{\epsilon \sigma}
\Psi_{E \sigma}^{\dag}|0> 
\label{eq:D} \\  
\phi_{4}  ( \epsilon )  &=& \sqt \sum_{\sigma}
\Psi_{\epsilon \sigma} f_{\sigma}^{\dag} |0> 
\label{eq:E} \\
\phi_{5} ( \epsilon_{1} , \epsilon_{2} ) &=& \sqt \sum_{\sigma}
\Psi_{\epsilon_{1} \sigma} \Psi_{\epsilon_{2} - \sigma}
c_{0  - \sigma}^{\dag} f_{\sigma}^{\dag} |0> 
\label{eq:F1} \\
\phi_{6} ( \epsilon_{1} , \epsilon_{2} ) &=& \sqt \sum_{\sigma}
\Psi_{\epsilon_{1} \sigma} \Psi_{\epsilon_{2} \sigma}
c_{0  \sigma}^{\dag} f_{\sigma}^{\dag} |0> 
\label{eq:F2} \\
\phi_{7} ( \epsilon_{1} , \epsilon_{2} , \epsilon_{3} ) &=&
\sqt \sum_{\sigma}\Psi_{\epsilon_{1} \sigma} \Psi_{\epsilon_{2} \sigma}
\Psi_{\epsilon_{3} - \sigma} c_{0 \sigma}^{\dag} c_{0 - \sigma}^{\dag}
f_{\sigma}^{\dag}|0> 
\label{eq:G} \\
\phi_{8} ( \epsilon_{1} , \epsilon_{2}, E ) &=& \sqt \sum_{\sigma}
\Psi_{\epsilon_{1} \sigma} \Psi_{\epsilon_{2} - \sigma}
\Psi_{E - \sigma}^{\dag} f_{\sigma}^{\dag} |0> 
\label{eq:H1} \\
\phi_{9} ( \epsilon_{1} , \epsilon_{2}, E ) &=& \sqt \sum_{\sigma} 
\Psi_{\epsilon_{1} \sigma} \Psi_{\epsilon_{2} \sigma}
\Psi_{E \sigma}^{\dag} f_{\sigma}^{\dag} |0>  \ .
\label{eq:H2} 
\end{eqnarray}

Beneath is the basis of the magnetic state, Eq. ( \ref{eq:psi-triplet} ): 
\begin{eqnarray}
\Phi_{1} ( \epsilon ) &=& \Psi_{\epsilon \da} c_{0 \ua}^{\dag} |0> 
\label{eq:Am} \\
\Phi_{2} ( \epsilon_{1} , \epsilon_{2} ) &=&
\Psi_{\epsilon_{1} \da} \Psi_{\epsilon_{2} \da}
c_{0 \da}^{\dag} c_{0 \ua}^{\dag} |0> 
\label{eq:Bm} \\
\Phi_{3}( \epsilon , E )  &=& \Psi_{\epsilon \da}
\Psi_{E \ua}^{\dag}|0> 
\label{eq:Dm} \\  
\Phi_{4}  ( \epsilon ) &=&
\Psi_{\epsilon \da} f_{\ua}^{\dag} |0> 
\label{eq:Em} \\
\Phi_{5} ( \epsilon_{1} , \epsilon_{2} ) &=&
\Psi_{\epsilon_{1} \da} \Psi_{\epsilon_{2} \da}
c_{0  \da}^{\dag} f_{\ua}^{\dag} |0> 
\label{eq:F1m} \\
\Phi_{6} ( \epsilon_{1} , \epsilon_{2} ) &=&
\Psi_{\epsilon_{1} \da} \Psi_{\epsilon_{2} \ua}
c_{0  \ua}^{\dag} f_{\ua}^{\dag} |0> 
\label{eq:F2m} \\
\Phi_{7} ( \epsilon_{1} , \epsilon_{2} ) &=&
\Psi_{\epsilon_{1} \da} \Psi_{\epsilon_{2} \da}
c_{0  \ua}^{\dag} f_{\da}^{\dag} |0> 
\label{eq:F3m} \\
\Phi_{8} ( \epsilon_{1} , \epsilon_{2} , \epsilon_{3} ) &=&
\Psi_{\epsilon_{1} \da} \Psi_{\epsilon_{2} \ua}
\Psi_{\epsilon_{3} \da} c_{0 \da}^{\dag} c_{0 \ua}^{\dag}
f_{\ua}^{\dag}|0> 
\label{eq:G1m} \\
\Phi_{9} ( \epsilon_{1} , \epsilon_{2} , \epsilon_{3} ) &=&
\Psi_{\epsilon_{1} \da} \Psi_{\epsilon_{2} \da}
\Psi_{\epsilon_{3} \da} c_{0 \da}^{\dag} c_{0 \ua}^{\dag}
f_{\da}^{\dag}|0> 
\label{eq:G3m} \\
\Phi_{10} ( \epsilon_{1} , \epsilon_{2}, E) &=&
\Psi_{\epsilon_{1} \da} \Psi_{\epsilon_{2} \da}
\Psi_{E \da}^{\dag} f_{\ua}^{\dag}|0> 
\label{eq:H1m} \\
\Phi_{11} ( \epsilon_{1} , \epsilon_{2}, E) &=&
\Psi_{\epsilon_{1} \da} \Psi_{\epsilon_{2} \ua}
\Psi_{E \ua}^{\dag} f_{\ua}^{\dag}|0> \label{eq:H2m} \\
\Phi_{12} ( \epsilon_{1} , \epsilon_{2} , E )  &=&
\Psi_{\epsilon_{1} \da} \Psi_{\epsilon_{2} \da}
\Psi_{E \ua}^{\dag} f_{\da}^{\dag}|0>  \ . 
\label{eq:H3m} 
\end{eqnarray}
\appendix{Integral equations}
\label{sec-intgrleq}
The equations for the functions $r_{4}, \cdots , r_{9}$ are 
\begin{eqnarray}
r_{4} ( \epsilon ) &=& \left[ \ V r_{1} ( \epsilon ) \right. \nonumber \\ 
&+&
\int_{-D}^{0} d \epsilon_{1} {\cal V} ( \epsilon_{1})
\left[ r_{5} ( \epsilon , \epsilon_{1} ) \right. \nonumber \\ 
&+& \left. \left.
\tilde{r}_{6} ( \epsilon , \epsilon_{1} ) \right] \right] /  
( \Delta E_{\em{s}} + \epsilon - \epsilon_{f}) \label{eq:ve} \\
&~& \nonumber \\
r_{5} ( \epsilon_{1} , \epsilon_{2} ) &=& \left[ \ V \tilde{r}_{2}
( \epsilon_{1} , \epsilon_{2} ) \right. \nonumber \\
&+& {\cal V} ( \epsilon_{2} ) 
r_{4} ( \epsilon_{1} )
+ \int_{-D}^{0} d \epsilon_{3} {\cal V} 
( \epsilon_{3} ) \tilde{r}_{7} ( \epsilon_{1} ,
\epsilon_{3} , \epsilon_{2} )  \nonumber \\
&+& \left. \int_{0}^{D} dE {\cal V} (E) r_{8} 
( \epsilon_{1} , \epsilon_{2} , E )
\right] \nonumber \\
&/& (  \Delta E_{\em{s}} + \epsilon_{1} + \epsilon_{2} + \mu 
- \epsilon_{f} ) \label{eq:vf1} \\
&~& \nonumber \\
\tilde{r}_{6} ( \epsilon_{1} , \epsilon_{2} ) &=& \left[ {\cal V} 
( \epsilon_{1} ) 
r_{4} ( \epsilon_{2})
+ {\cal V} ( \epsilon_{2} ) r_{4} ( \epsilon_{1} ) \right. \nonumber \\
&+& \int_{-D}^{0} d \epsilon_{3} {\cal V}
( \epsilon_{3} ) \tilde{r}_{4} ( \epsilon_{1} ,
\epsilon_{2} , \epsilon_{3} )  \nonumber \\
&+& \left. \int_{0}^{D} dE {\cal V}(E) \tilde{r}_{9}  
( \epsilon_{1} , \epsilon_{2} , E )
\right] \nonumber \\ 
&/&   (  \Delta E_{\em{s}} + \epsilon_{1} + \epsilon_{2} + \mu 
- \epsilon_{f} ) \label{eq:vf2} \\
&~& \nonumber \\
\tilde{r}_{7} ( \epsilon_{1} , \epsilon_{2} , \epsilon_{3} ) &=& 
\left[ {\cal V}( \epsilon_{1} )  r_{5} ( \epsilon_{1} , \epsilon_{2} ) + 
{\cal V} ( \epsilon_{2} ) r_{5} ( \epsilon_{1} , \epsilon_{3} ) \right. 
\nonumber \\
&+& \left. {\cal V}( \epsilon_{3} ) 
\tilde{r}_{6} ( \epsilon_{1} , \epsilon_{2} ) \right] \nonumber \\
&/& 
(  \Delta E_{\em{s}} + \epsilon_{1} + \epsilon_{2} + \epsilon_{3} - 
\epsilon_{f} ) \label{eq:vg} \\
&~& \nonumber \\
r_{8} ( \epsilon_{1} , \epsilon_{2} , E) &=& {\cal V}(E) 
r_{5} ( \epsilon_{1} , \epsilon_{2} ) \nonumber \\
&/& \ ( \Delta E_{\em{s}}
  + \epsilon_{1} + \epsilon_{2} - E - \epsilon_{f} ) \label{eq:vh1} \\
 &~& \nonumber \\
\tilde{r}_{9} ( \epsilon_{1} , \epsilon_{2} , E) &=& {\cal V}(E) 
\tilde{r}_{6} ( \epsilon_{1} , \epsilon_{2} ) 
\nonumber \\
&/& ( \Delta E_{\em{s}} + \epsilon_{1} + \epsilon_{2} - E - 
\epsilon_{f} )  . \label{eq:vh2} 
\end{eqnarray}
where
\begin{eqnarray}
\tilde{r}_{2} ( \epsilon_{1} , \epsilon_{2} ) &=& 
\Theta( \epsilon_{1} - \epsilon_{2} ) r_{2} ( \epsilon_{1} , \epsilon_{2} ) 
\nonumber \\ 
&+&
\Theta( \epsilon_{2} - \epsilon_{1} ) r_{2} ( \epsilon_{2} , \epsilon_{1} ) \\
&~& \nonumber \\
\tilde{r}_{6} ( \epsilon_{1} , \epsilon_{2} ) &=& 
\Theta( \epsilon_{1} - \epsilon_{2} ) r_{6} ( \epsilon_{1} , \epsilon_{2} ) 
\nonumber \\
&+&
\Theta( \epsilon_{2} - \epsilon_{1} ) r_{6} ( \epsilon_{2} , \epsilon_{1} ) \\
&~& \nonumber \\
\tilde{r}_{7} ( \epsilon_{1} , \epsilon_{2} , \epsilon_{3} ) &=& 
\Theta( \epsilon_{1} - \epsilon_{2} ) 
r_{7} ( \epsilon_{1} , \epsilon_{2} , \epsilon_{3} ) \nonumber \\
&+&
\Theta ( \epsilon_{2} - \epsilon_{1} ) 
r_{7} ( \epsilon_{2} , \epsilon_{1} , \epsilon_{3} ) \\
&~& \nonumber \\
\tilde{r}_{9} ( \epsilon_{1} , \epsilon_{2} ,E ) &=& 
\Theta( \epsilon_{1} - \epsilon_{2} ) 
r_{9} ( \epsilon_{1} , \epsilon_{2} ,E) \nonumber \\
&+&
\Theta( \epsilon_{2} - \epsilon_{1} )  r_{9} ( \epsilon_{2} , \epsilon_{1} ,E)
 . \end{eqnarray}
\vspace{1cm}

The equations for the functions $R_{1}, \cdots , R_{12}$ are
\begin{eqnarray}
R_{1} ( \epsilon ) &=& \left[ \ V R_{4} ( \epsilon ) \right. \nonumber \\
&+&
\int_{-D}^{0} d \epsilon_{1} {\cal V} ( \epsilon_{1} ) 
\tilde{R}_{2} ( \epsilon , \epsilon_{1} )  \nonumber \\
&+& \left. \int_{0}^{D} dE {\cal V}(E) 
d ( \epsilon , E ) \right] \nonumber \\
&/& 
\ ( \Delta E_{\em{t}} + \epsilon +\mu )
 \label{eq:vam} \\
&~& \nonumber \\
\tilde{R}_{2} ( \epsilon_{1} , \epsilon_{2} ) &=& \left[ 
{\cal V} ( \epsilon_{1} ) 
R_{1} ( \epsilon_{2} ) + {\cal V} ( \epsilon_{2}) R_{1} ( \epsilon_{1} ) 
\right. 
\nonumber \\
&+& \left. V \left[ \tilde{R}_{5} ( \epsilon_{1} ,\epsilon_{2} )
+ \tilde{R}_{7} ( \epsilon_{1} , \epsilon_{2} ) \right]  \right] \nonumber \\
&/& \ 
( \Delta E_{\em{t}} + \epsilon_{1} + \epsilon_{2} )
 \label{eq:vbm} \\
&~& \nonumber \\
R_{3} ( \epsilon , E ) &=& {\cal V} (E) R_{1} ( \epsilon ) 
\ / \ ( \Delta E_{\em{t}} + \epsilon - E) \label{eq:vdm} \\
&~& \nonumber \\
R_{4} ( \epsilon ) &=& \left[ \ V R_{1} ( \epsilon ) \right. \nonumber \\
&+&
\int_{-D}^{0} d \epsilon_{1} {\cal V} ( \epsilon_{1}) 
\left[ \tilde{R}_{5} ( \epsilon , \epsilon_{1} ) \right. \nonumber \\
&+& 
\left. \left. R_{6} ( \epsilon , \epsilon_{1} ) \right] \right] \nonumber \\ 
&/& \  ( \Delta E_{\em{t}} + \epsilon - \epsilon_{f} )
 \label{eq:vem} \\
&~& \nonumber \\
\tilde{R}_{5} ( \epsilon_{1} , \epsilon_{2} ) &=& 
\left[ \ V \tilde{R}_{2}
( \epsilon_{1} , \epsilon_{2} ) \right. \nonumber \\
&+& {\cal V} ( \epsilon_{2} ) 
R_{4} ( \epsilon_{1} )
+ {\cal V} (\epsilon_{1}) R_{4} (\epsilon_{2})  \nonumber  \\
&+& 
\int_{-D}^{0} d \epsilon_{3} {\cal V} ( \epsilon_{3} ) \tilde{R}_{8} 
( \epsilon_{1} , \epsilon_{3} , \epsilon_{2} ) \nonumber \\
&+& \left.  \int_{0}^{D} dE {\cal V}(E) \tilde{R}_{10} 
( \epsilon_{1} , \epsilon_{2} , E )
\right] \nonumber \\ 
&/& \  
( \Delta E_{\em{t}} + \epsilon_{1} + \epsilon_{2} + \mu - \epsilon_{f} )
\label{eq:vf1} \\
&~& \nonumber \\
R_{6} ( \epsilon_{1} , \epsilon_{2} ) &=& \left[
 {\cal V}( \epsilon_{2} ) R_{4} ( \epsilon_{1} ) \right. \nonumber \\
&+& \int_{-D}^{0} d \epsilon_{3} 
{\cal V}( \epsilon_{3} ) \tilde{R}_{8} 
( \epsilon_{1} ,\epsilon_{2} , \epsilon_{3} )  \nonumber \\
&+& \left. \int_{0}^{D} dE {\cal V}(E) R_{11} 
( \epsilon_{1} , \epsilon_{2} , E )
\right] \nonumber \\  
&/& \ ( \Delta E_{\em{t}} + \epsilon_{1} + \epsilon_{2} + \mu - \epsilon_{f} )
 \label{eq:vf2m} \\
&~& \nonumber \\
\tilde{R}_{7} ( \epsilon_{1} , \epsilon_{2} ) &=& 
\left[ \ V \tilde{R}_{2}
( \epsilon_{1} , \epsilon_{2} ) \right. \nonumber \\ 
&+&
\int_{-D}^{0} d \epsilon_{3} {\cal V}( \epsilon_{3} ) \tilde{R}_{9} 
( \epsilon_{1} , \epsilon_{2} , \epsilon_{3} )  \nonumber \\ 
&+& \left.  \int_{0}^{D} dE {\cal V}(E) \tilde{R}_{12} 
( \epsilon_{1} , \epsilon_{2} , E )
\right] \nonumber \\ 
&/& \ 
( \Delta E_{\em{t}} + \epsilon_{1} + \epsilon_{2} + \mu - \epsilon_{f} ) 
\label{eq:vf3m} \\
&~& \nonumber \\
\tilde{R}_{8} ( \epsilon_{1} , \epsilon_{2} , \epsilon_{3} ) &=& 
\left[ {\cal V}( \epsilon_{1} )  R_{6} ( \epsilon_{2} , \epsilon_{3} ) 
\right. \nonumber \\ 
&+&
{\cal V} ( \epsilon_{3} ) R_{6} ( \epsilon_{1} , \epsilon_{2} ) 
\nonumber \\
&+& \left. {\cal V}( \epsilon_{2} ) 
\tilde{R}_{5} ( \epsilon_{1} , \epsilon_{3} ) \right] \nonumber \\ 
&/& \ 
( \Delta E_{\em{t}} + \epsilon_{1} + \epsilon_{2} + 
\epsilon_{3} - \epsilon_{f} )
 \label{eq:vg1m} \\
&~& \nonumber \\
\tilde{R}_{9} ( \epsilon_{1} , \epsilon_{2} , \epsilon_{3} ) &=& 
\left[ {\cal V} ( \epsilon_{1} )  \tilde{R}_{7} 
( \epsilon_{2} , \epsilon_{3} ) \right. \nonumber \\ 
&+& 
{\cal V} ( \epsilon_{2} ) \tilde{R}_{7} 
( \epsilon_{1} , \epsilon_{3} ) 
\nonumber \\
&+& \left. {\cal V}( \epsilon_{3} ) 
\tilde{R}_{7} ( \epsilon_{1} , \epsilon_{2} ) \right] \nonumber \\
&/& \ 
( \Delta E_{\em{t}} + \epsilon_{1} + \epsilon_{2} + 
\epsilon_{3} - \epsilon_{f} )
\label{eq:vg3m} \\
&~& \nonumber \\
\tilde{R}_{10} ( \epsilon_{1} , \epsilon_{2} , E) &=& {\cal V}(E) 
\tilde{R}_{5} ( \epsilon_{1} , \epsilon_{2} ) 
\nonumber \\ 
&/& \ ( \Delta E_{\em{t}} + \epsilon_{1} +
\epsilon_{2} - E - \epsilon_{f} ) 
\label{eq:vh1m} \\
&~& \nonumber \\
R_{11} ( \epsilon_{1} , \epsilon_{2} , E) &=& {\cal V}(E) 
R_{6} ( \epsilon_{1} , \epsilon_{2} ) 
\nonumber \\ 
&/&  \ (  \Delta E_{\em{t}} + \epsilon_{1} +
\epsilon_{2} - E - \epsilon_{f} ) 
\label{eq:vh2m} \\
&~& \nonumber \\
\tilde{R}_{12} ( \epsilon_{1} , \epsilon_{2} , E) &=& {\cal V}(E) 
\tilde{R}_{7} ( \epsilon_{1} , \epsilon_{2} ) 
\nonumber \\ 
&/& \ ( \Delta E_{\em{t}} + \epsilon_{1} +
\epsilon_{2} - E - \epsilon_{f} ) \ . 
\label{eq:vh3m} 
\end{eqnarray}
\vspace{0.5cm}

where
\begin{eqnarray}
\tilde{R}_{2} ( \epsilon_{1} , \epsilon_{2} ) &=& 
\Theta ( \epsilon_{1} - \epsilon_{2} ) R_{2} ( \epsilon_{1} , \epsilon_{2} ) 
\nonumber \\
&+&
\Theta ( \epsilon_{2} - \epsilon_{1} ) R_{2} ( \epsilon_{2} , \epsilon_{1} ) \\
&~& \nonumber \\
\tilde{R}_{5} ( \epsilon_{1} , \epsilon_{2} ) &=& 
\Theta ( \epsilon_{1} - \epsilon_{2} ) 
R_{5} ( \epsilon_{1} , \epsilon_{2} ) \nonumber \\ &+&
\Theta ( \epsilon_{2} - \epsilon_{1} ) 
R_{5} ( \epsilon_{2} , \epsilon_{1} ) \\
&~& \nonumber \\
\tilde{R}_{7} ( \epsilon_{1} , \epsilon_{2} ) &=& 
\Theta ( \epsilon_{1} - \epsilon_{2} ) 
R_{7} ( \epsilon_{1} , \epsilon_{2} ) \nonumber \\ &+&
\Theta ( \epsilon_{2} - \epsilon_{1} ) 
R_{7} ( \epsilon_{2} , \epsilon_{1} ) \\
&~& \nonumber \\
\tilde{R}_{8} ( \epsilon_{1} , \epsilon_{2} , \epsilon_{3} ) &=& 
\Theta ( \epsilon_{1} - \epsilon_{3} ) 
R_{8} ( \epsilon_{1} , \epsilon_{2} , \epsilon_{3} ) \nonumber \\
&+&
\Theta ( \epsilon_{2} - \epsilon_{1} ) 
R_{8} ( \epsilon_{3} , \epsilon_{1} , \epsilon_{1} ) \\
&~& \nonumber \\
\tilde{R}_{9} ( \epsilon_{1} , \epsilon_{2} , \epsilon_{3} ) &=& 
\Theta ( \epsilon_{1} - \epsilon_{2} ) 
\Theta ( \epsilon_{2} - \epsilon_{3} )  
R_{9} ( \epsilon_{1} , \epsilon_{2} , \epsilon_{3} ) \nonumber \\
&+&
\Theta ( \epsilon_{1} - \epsilon_{3} )
\Theta ( \epsilon_{3} - \epsilon_{2} )
R_{9} ( \epsilon_{1} , \epsilon_{3} , \epsilon_{2} ) \nonumber \\
&+& \Theta ( \epsilon_{2} - \epsilon_{1} ) 
\Theta ( \epsilon_{1} - \epsilon_{3} )  
R_{9} ( \epsilon_{2} , \epsilon_{1} , \epsilon_{3} ) \nonumber \\ 
&+&
\Theta ( \epsilon_{2} - \epsilon_{3} ) 
\Theta ( \epsilon_{3} - \epsilon_{1} )  
R_{9} ( \epsilon_{2} , \epsilon_{3} , \epsilon_{1} ) \nonumber \\
&+& \Theta ( \epsilon_{3} - \epsilon_{1} ) 
\Theta ( \epsilon_{1} - \epsilon_{2} )  
R_{9} ( \epsilon_{3} , \epsilon_{1} , \epsilon_{2} ) \nonumber \\
&+&
\Theta ( \epsilon_{3} - \epsilon_{2} ) 
\Theta ( \epsilon_{2} - \epsilon_{1} )  
R_{9} ( \epsilon_{3} , \epsilon_{2} , \epsilon_{1} )   \\
&~& \nonumber \\
\tilde{R}_{10} ( \epsilon_{1} , \epsilon_{2} ,E ) &=& 
\Theta ( \epsilon_{1} - \epsilon_{2} ) 
R_{10} ( \epsilon_{1} , \epsilon_{2} ,E) \nonumber \\
&+&
\Theta ( \epsilon_{2} - \epsilon_{1} ) 
R_{10} ( \epsilon_{2} , \epsilon_{1} ,E) \\
&~& \nonumber \\
\tilde{R}_{12} ( \epsilon_{1} , \epsilon_{2} ,E ) &=& 
\Theta ( \epsilon_{1} - \epsilon_{2} ) 
R_{12} ( \epsilon_{1} , \epsilon_{2} ,E) \nonumber \\
&+&
\Theta ( \epsilon_{2} - \epsilon_{1} ) 
R_{12} ( \epsilon_{2} , \epsilon_{1} ,E) \ . 
\end{eqnarray}
and $\Delta E_{\em{t}}$ is the energy difference 
between the lowest magnetic state energy and the energy of the filled Fermi 
sea of the effective conduction band, Eq.(\ref{eq:H_Psi}) .


\appendix{Numerics}
\label{sec-iterations}
A good illustration of the method is provided by  the Anderson 
impurity model with a finite $U$ \cite{GS2,footnote3}. In our case, putting 
in Eqs.(\ref{eq:intgrleq1}), (\ref{eq:intgrleq2})
$V = 0$, $\mu = U/2$ and taking ${\cal V}(\epsilon)$ from Figure 
\ref{fig:DOS_U} we obtain the variational principle equations for the 
singlet state of the symmetric Anderson model which represents the original 
Hubbard host, Eq. (\ref{eq:H_h}). We fix a value of $\Delta E$ and choose the
following functions for starting the iterations: $r_2 = r_3 = 0$ and $r_1(
\epsilon) = \sqrt{2}{\cal V}(\epsilon)/(\Delta E + \epsilon + U/2 )$. Then,
instead of the normalization we keep $r_0 = 1$ on all stages of iterations.
The convergence is reached in a few steps and the minimal fixed point
of the function $f$, Eq. (\ref{eq:fixedpoint_f}), is found. The same method 
was applied to find $\Delta E_{\em{t}}$ for the symmetric Anderson model. 
We define $\Delta E_{\em{t}} - \Delta E_{\em{s}} = T_{K}^{AM}$ and 
find that $T_{K}^{AM}/D$ = 0.081, 0.025, 0.011, 0.003 for $U/D$ = 1.0, 
2.0, 2.5 ,3.0 respectively. We estimate the accuracy of above calculations  
as $\pm 10^{-4}$. We may check the credibility of our fixed point iterations 
by comparing the above values of $T_{K}^{AM}$ with the width of the 
relevant central peak of the DOS in Figure \ref{fig:DOS_U}. This width has 
to be proportional to $T_{K}^{AM}$ \cite{LISAreview} and one can see
 that it is a fairly good agreement indeed between widths ratios and ratios 
of $T_{K}^{AM}$ for $U/D = 1,2,2.5$ while for the $U/D = 3$ the 
agreement is less good. The latter is expected because the relative error in 
this case much larger than in three others. The other check is provided by 
the expected exponential dependence of $T_{K}^{AM}/D$ upon $U$ 
\cite{TsvelickWiegmann}. The latter is confirmed by an almost linear 
dependence between the calculated $ln{T_{K}^{AM}}$ and $U$. 

In calculating the singlet ground state energy of the two-impurity Hamiltonian 
 $H_{\em{eff}}$ we choose as the
initial guess for the superposition in Eq.~(\ref{eq:singlet}) the following 
functions:
\begin{eqnarray}
r_{4}(\epsilon_{1},\epsilon_{2}) &=& r_{5}(\epsilon_{1},\epsilon_{2}) =
\delta (\epsilon_{1}) \delta (\epsilon_{2}) \nonumber \\
r_{i} &=& 0 \ \ \ \ i \neq 4,5  \  
\label{initial-singlet}
\end{eqnarray}
The reason for this choice is that in the limit 
$V, {\cal V} \rightarrow 0$ (i.e. no interaction between the 
superposition states) the states $\phi_{4} (0,0), \phi_{5} (0,0)$ are the 
states with lowest energies and therefore the only ones which are occupied .
The smaller $V$ the faster the convergence of the iterations .

For similar reasons the initial guess for the lowest-lying triplet 
superposition is 
\begin{eqnarray}
R_{5}(\epsilon_{1},\epsilon_{2}) &=& R_{6}(\epsilon_{1},\epsilon_{2}) =
\delta (\epsilon_{1}) \delta (\epsilon_{2}) \nonumber \\
r_{i} &=& 0 \ \ \ \ i \neq 4,5  \  
\label{initial-triplet}  
\end{eqnarray}      
An illustration of the iteration procedure for the singlet case is given in
Figure~\ref{fig:iteration} .
As it is known  (~\cite{VarmaYafet},~\cite{GS1},~\cite{GS2}) 
the variational wavefunction of the Anderson impurity model sharp varies
in the vicinity of zero energy (i.e. Fermi energy).     
Therefore in order to achieve sufficiently accurate integration
with a small number of mesh points
we used the dense mesh in the vicinity of zero:
\begin{equation}
\epsilon_{j} = e^{\frac{j}{n_{d}} \ln (1+g/D)} -1 \label{eq:logmesh} 
\end{equation}
where
\begin{eqnarray}
0 & \leq & j \leq n_{d} \\
0 & \leq & g/D \ll 1 \nonumber
\end{eqnarray} 
$g$ is the limit of the dense mesh and $n_{d}$ is the number of dense 
mesh points . 
In the range $g < E \leq D $  
the variational wavefunctions is usually smooth so we used a Gauss-Legendre
mesh . The best $g/D$ was found to be $10^{-3}$ . 

As a criterion for convergence of the iteration procedure we took the 
inequality
\begin{equation}
J(i) = |f^{i+1}( \Delta E) - f^{i}( \Delta E)| < c
\label{eq:ineq}
\end{equation}
where $f^{i}(\Delta E)$ is the value of the function $f(\Delta E)$, Eq. (\ref 
{eq:fixedpoint_f}), in the $i$-th iteration and $c$ is a small number. 
Surprisingly we found that the inequality 
\begin{equation}
J(i+2)/J(i+1) - J(i+1)/J(i) < c
\end{equation}
holds for $i$ which is much smaller as compared with a needed for the 
existence of the inequality in Eq. (\ref{eq:ineq}). 
So after obtaining a convergence of the factor
$ J(i+1)/J(i) $ we calculated  $f(\Delta E)$
by the geometric series sum rule
\begin{equation}
f(\Delta E) = \frac{f^{i}(\Delta E)}{1- J(i+1)/J(i)} \  .
\label{eq:geometric}
\end{equation}
The inequality $ J(i+1)/J(i) >1 $ indicates divergence of the
iterations .
%

\figure{%
A Bethe lattice with the impurity coupled to a lattice site.
\label{fig:Bethe-lattice}}

\figure{%
$2t(-ImG(\epsilon))$ versus $\epsilon/2t$ as calculated in \cite{?} for the 
Hubbard model on Bethe lattice with the infinite coordination, 
$d \rightarrow \infty$. U = 0;1;2;2.5;3;4.
\label{fig:DOS_U}}


\figure{%
Slave-boson and f-pseudofermions self-energies. The solid, dashed and wavy 
lines represent propagators for the $c_0$-electron, f-pseudofermion and slave
boson. The open circle denotes the bare hybridization interaction V while 
the filled square is two-particle vertex $\Gamma^{U}(12;34)$. a) The slave 
boson self-energy, $U=0$. b) The f-pseudofermion self-energy, $U=0$. c) and 
d) are additional contributions to self-energies a) and b) respectively for 
$U \neq 0$. %
\label{fig:bubbles}}

\figure{%
An illustration of states $\phi_i$ which composite the singlet 
variational state, Eq. (\ref{eq:singlet}). Explanations of $\phi$-states are 
in the text and in the Appendix \ref{sec-tables}. Solid lines connect states 
which are coupled by the ${\cal V}(\epsilon)$-interaction and dashed lines 
are for the hybridization interaction V ( see Eq. (\ref{eq:H_Psi})).%
\label{fig:boxes_s}}


\figure{%
$T_K$ as a function of $V^2$ for $U$ =1 and $\epsilon_f$ = - 0.3 
( upper curve ) and - 0.67 ( low curve ). All parameters in units of $D$. 
Curves are a guide to the eye.    
\label{fig:curves_U=1}}

\figure{%
The same as in Figure \ref{fig:curves_U=1} but for different values of $U$.
Upper curve: $U/D = 2$, lower curve:$U/D = 2.5$.
$\epsilon_f$ = - 0.3. Curves are a guide to the eye. 
\label{fig:curves_U=2}}

\figure{%
$T_K$ dependence on $U$ for$\epsilon_f$ = - 0.3 and $V^2$ = 0.01. The cross 
marks $T_K$ from Eq. (\ref{eq:T_K}). 
\label{fig:curve_U}}

\figure{%
An illustration of the iteration procedure for the singlet case.
\label{fig:iteration}}


\begin{thebibliography}{10}
\bibitem{footnote1}
We use here the notion of the Kondo effect as the generic one for all 
phenomena of a magnetic impurity in a metallic host.
\bibitem{footnote2}
The opposite effect was found in the paper: K. Itai and P. Fazekas, Phys. Rev
  B {\bf 54},  R752  (1996). Vertex corrections are not taking into account
in this paper and that may be the reason of the discrepancy.
\bibitem{footnote3}
We applied this iteration scheme to the finite-U Anderson impurity of ref.
\cite{GS2} and easily obtained state occupancies and Kondo temperatures cited
in ref.\cite{GS2}.
\bibitem{footnote4}
The meaninig of the relevant magnetic states in the context of the variational 
approach to the Kondo effect is discussed in details in reference \cite{GS2}. 
These are states which contribute to the Anderson impurity magnetic 
susceptibility where the Kondo scale appears. There are many other magnetic 
states with much lower energies but they are not connected by the impurity 
Zeeman interaction with the singlet ground state \cite{GS2}.
\bibitem{footnote5}
We note that in the
non-interacting case the Kondo temperature as a function of the Anderson width
$\Gamma$ may have a maximum. This happens in a transitional region from the 
Kondo regime ( where $T_K$ increases exponentialy with $\Gamma$ ) to 
the mixed valence one when a scale becomes of the order of 
$\epsilon_{f}^2/\Gamma$ ( for $\Gamma \stackrel {<} {\sim} |\epsilon_f|$ ) and 
eventually transforms 
to $\sim \Gamma$ ( for $\Gamma > |\epsilon_f|$ ) \cite{TsvelickWiegmann}.
\end{thebibliography}
\end{document}